# SUPERCONDUCTIVITY IN ACTINIDE MATERIALS

J. D. Thompson[1], J. L. Sarrao[1], N. J. Curro[1], E. D. Bauer[1], L. A. Morales[1], F. Wastin[2], J. Rebizant[2], J. C. Griveau[2], P. Boulet[2], E. Colineau[2] and G. H. Lander[2]

[1]Los Alamos National Laboratory, Los Alamos, NM 87545 USA
[2]European Commission, JRC, Institute for Transuranium Elements, Post Box 2340, 76125 Karlsruhe, Germany

## 1 INTRODUCTION

More than four decades after the first discovery of superconductivity in elemental mercury, Bardeen, Cooper and Schrieffer showed theoretically that superconductivity arises when conduction-band electrons form pairs with precisely opposite spin and momentum due to an attractive interaction mediated by lattice vibrations (phonons).[1] With zero net spin and angular momentum, these Cooper pairs condense into a macroscopic quantum state that is separated energetically from all unpaired electrons by a finite gap $\Delta$, which is proportional to the superconducting transition temperature $T_c$. Though unable to predict what materials might be superconducting, this theory established conditions favourable for superconductivity, namely that there be a high density of electronic states at the Fermi energy ($N(E_F)$) and phonons should be sufficiently soft to create Cooper pairs effectively. Relative to s, p metals, transition metals have a large density of electronic states, and, consequently, these concepts provided a natural explanation for why superconductivity prevailed in d-electron metals.[2]

The same conditions for the appearance of superconductivity apply to actinides that are d-electron-like. Tetravalent Th is one such example and is superconducting below 1.39 K.[3] As with d-electron metals, certain crystal structures also tend to favour superconductivity in actinide-based metals. A simple rationale is based on Hill's criterion for the overlap of 5f wave functions.[4] For actinide-actinide spacings less than ~0.35 (~0.34) nm for U (Pu), wave functions of 5f electrons from adjacent actinides overlap and create a narrow band of itinerant electrons, with predominantly 5f symmetry, that crosses the Fermi energy. Qualitatively, the width of this electronic band is inversely proportional to $N(E_F)$, i.e., narrower bands correspond to a higher density of electronic states at $E_F$. Crystal structures that impose nearest actinide distances below Hill's limit, then, are a necessary but not sufficient condition for phonon-mediated superconductivity. For sufficiently narrow bands, Coulomb interactions between electrons become increasingly important. In the BCS theory, these repulsive Coulomb interactions compete with the attractive interaction provided by phonons and reduce the superconducting transition temperature that otherwise would have been possible.[2]

In metals with actinide-actinide spacings beyond Hill's limit, 5f wave functions do not overlap directly, the 5f electrons are quasi-localized in states below $E_F$, and, consequently, they should not participate in superconductivity. Nevertheless, there are several examples of materials that, by this reasoning, should not superconduct, but they do; furthermore, the 5f electrons play an essential role in the superconductivity.[5] These 'anomalous' superconductors belong to a family called strongly correlated or heavy-fermion materials in which a very narrow band, of order 1-10 meV wide, forms at low temperatures through hybridisation of 5f and ligand wave functions. Precisely how the physics of these heavy-fermion compounds should be understood theoretically remains a major challenge for condensed matter physics. At least qualitatively, though, some things are known.[6] Above a few to tens of Kelvins, the magnetic susceptibility of these materials follows a Curie or Curie-Weiss temperature dependence with an effective moment close to that expected if the f-electrons were fully localized. At much lower temperatures, the magnetic susceptibility tends to a large constant value that is typical of a massively enhanced Pauli-type susceptibility, which is proportional to $N(E_F)$. In this same low temperature regime, the electronic coefficient of specific heat $\gamma$, also proportional to $N(E_F)$, grows correspondingly. At extremely low temperatures, direct measurements of the effective mass $m^*$ of itinerant electrons at $E_F$ find that $m^*$ reaches 100-1000 times the mass of a free electron, consistent with values of $m^*$ implied by the massively enhanced values of $\gamma$. These same heavy charge carriers participate in superconductivity as evidenced by a jump in specific heat at $T_c$ that follows the BSC prediction $\Delta C/\gamma T_c \approx 1.5$.[5,6] Excluding their superconductivity, properties of heavy-fermion materials are qualitatively similar to behaviours exhibited by an isolated local moment that interacts antiferromagnetically with a broad band of conduction electrons through the Kondo effect to create a resonance in the density of states near $E_F$.[6] Associated with this resonance is a large $N(E_F)$ with character dominated by the symmetry of the local moment, i.e., 5f character in the case of actinides. This simple picture, however, presents a dilemma. Introduction of Kondo impurities into conventional superconductors rapidly suppresses $T_c$ to zero because of the pair-breaking effect of the local moments,[7] but heavy-fermion superconductors are built from a periodic array of approximately $10^{23}$ Kondo impurities, e.g., Ce, U and Pu. The mere existence of heavy-fermion superconductivity suggests that the conventional mechanism of phonon-mediated superconductivity is inappropriate and that alternative mechanisms should be considered.

One alternative is that Cooper pairing is mediated by spin fluctuations,[5,8] which, like phonons, are bosonic excitations. Cooper pairs formed by the exchange of spin fluctuations do not necessarily have zero net spin and momentum that is required in conventional superconductivity.[8] Because paired states with spin $S \geq 0$ and angular momentum $L \geq 0$ are allowed, the superconducting energy gap $\Delta$ goes to zero on certain parts of the Fermi surface. Figure 1 compares an unconventional superconducting gap with $S=0$ and $L=2$ to a conventional gap, which is finite over the entire Fermi surface. One consequence of gap nodes, where $\Delta=0$, is that $N(E_F)$ is non-zero as temperature goes to zero, and physical properties that depend on $N(E_F)$ will grow as a power-law in temperature for $T \ll T_c$, in stark contrast to a thermally activated temperature dependence in a conventional superconductor.[9] In the following, we take a working definition of an unconventional superconductor as one in which the uniform magnetic susceptibility is local-moment-like, the actinide-actinide spacing is close to or beyond the Hill limit, and power-law temperature variations appear below $T_c$ in physical properties that depend on $N(E_F)$.

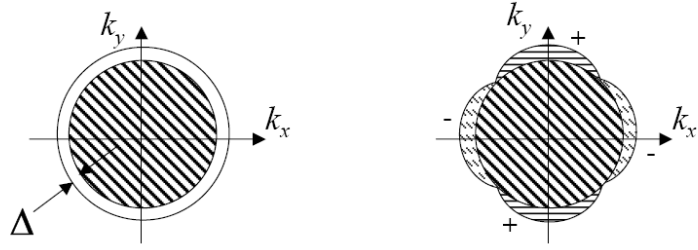

**Figure 1** *Schematic representation of the energy gap of a conventional superconductor with finite gap $\Delta$ over the entire Fermi surface (left) and of an unconventional superconductor with gap nodes (right).*

## 2 DISCUSSION

Table 1 gives, to our knowledge, a complete list of all actinide superconductors. Uranium-based superconductors known prior to 1989 (those listed before UGe$_2$ in Table 1) have been discussed already in excellent reviews in which arguments are presented that some of these superconductors, e.g., $UPt_3$, $UBe_{13}$, are unconventional.[5,30] Since that time, several new actinide-based superconductors have been discovered, including those based on Pu, and these are subjects of brief discussion in the next three sections.

**Table 1** *Actinide superconductors, with their transition temperature $T_c$ and nearest f-f spacing d.*

| Material | $T_c$ (K) | d (nm) | Material | $T_c$ (K) | d (nm) |
|---|---|---|---|---|---|
| α-U | <0.1[10,a] | 0.31 | U$_2$PtC$_2$ | 1.47[20] | 0.35 |
| β-U | 0.75-0.85[11,b] | 0.31 | UAl$_2$Si$_2$ | 1.34[16] | 0.41 |
| γ-U | 1.85-2.07[12,c] | 0.29 | UAl$_2$Ge$_2$ | 1.60[16] | 0.42 |
| UCo | 1.22[13] | 0.27/0.36[c] | UGa$_2$Ge$_2$ | 0.87[16] | 0.42 |
| U$_6$Fe | 3.78[13] | 0.32 | URu$_2$Si$_2$ | 1.5[21,f] | 0.41 |
| U$_6$Mn | 2.31[13] | 0.32 | UGe$_2$ | 0.4[22,g] | 0.38 |
| U$_6$Co | 2.33[13] | 0.32 | UIr | 0.14[23,h] | 0.33-0.38[g] |
| U$_6$Ni | 0.33[14] | 0.32 | UPd$_2$Al$_3$ | 1.9[24,i] | 0.40 |
| UPt$_3$ | 0.54[15,d] | 0.41 | UNi$_2$Al$_3$ | 1.0[25,j] | 0.40 |
| URu$_3$ | 0.145[16] | 0.40 | URhGe | 0.25[26,kj] | 0.35 |
| UBe$_{13}$ | 0.9[17] | 0.51 | PuCoGa$_5$ | 18.5[27] | 0.42 |
| U$_3$Ir | 1.3[18] | 0.40 | PuRhGa$_5$ | 8.7[28] | 0.43 |
| U$_5$Ge$_3$ | 0.99[19,e] | 0.29/0.36[c] | Am | 0.79[29,l] | 0.30 |

[a]$T_c$ exceeds 2 K under pressure; [b]$T_c$ depends on elements included to stabilize crystal structure; [c]Two inequivalent U sites; [d]Superconductivity coexists with weak magnetism ($T_N$=5 K); [e]Superconductivity controversial; [f]Superconductivity coexists with weak magnetism ($T_N$=17.5K); [g]Superconductivity pressure induced at 1<$P_c$<1.6 GPa and coexists with weak itinerant ferromagnetism; [h]Four inequivalent U sites and superconductivity pressure induced at $P_c \approx$ 2.6 GPa, near the boundary of weak itinerant ferromagnetism; [i]Superconductivity coexists with strong antiferromagnetism ($T_N$=14 K); [j]Superconductivity coexists with weak magnetism ($T_N$=4.6 K); [k]Superconductivity coexists with weak itinerant ferromagnetism ($T_C$=9.5 K); [l]$T_c$ exceeds 2 K under pressure, as discussed in the text.

## 2.1 Uranium Superconductors

In each of the U-based superconductors, $UGe_2$, UIr, URhGe, $UPd_2Al_3$ and $UNi_2Al_3$, unconventional superconductivity, by our working definition, exists simultaneously with some form of magnetism that derives from their 5f electrons. The first three compounds are 5f-band ferromagnets; whereas, the hexagonal U123 materials are antiferromagnetic with Néel temperatures well above $T_c$. The normal and superconducting state properties of $UGe_2$, URhGe and $UPd_2Al_3$ are summarized in [31], which includes an extensive bibliography of the literature on these materials.

There is little doubt that U's 5f electrons are responsible for the magnetism in these materials, but a crucial question is whether these same 5f electrons participate in superconductivity. Even in conventional superconductors, local moment antiferromagnetism and superconductivity can coexist, provided that antiferromagnetic order of localized f electrons does not couple to d-electrons responsible for superconductivity.[32] Like $UPt_3$[33] and $URu_2Si_2$,[34] $UPd_2Al_3$ is an example of superconductivity coexisting with antiferromagnetism, but in $UPd_2Al_3$ the relationship between superconductivity and magnetic order has been made especially clear and may bear more broadly on understanding other actinide superconductors. Neutron-diffraction studies of $UPd_2Al_3$ find[35] an atomic-like staggered moment (0.84 $\mu_B$) below $T_N$, as expected for localized 5f orbitals; however, the jump in specific heat at $T_c$, $\Delta C/\gamma T_c \approx 1.48$ with $\gamma = 115$ mJ/mol-$K^2$, is consistent with superconductivity developing out of a band of delocalized heavy-mass electrons.[36] This apparent dichotomy has led to the suggestion[36] that 5f electrons in $UPd_2Al_3$ assume dual roles, with two of the three 5f electrons being localized (magnetic)[37] and the other being itinerant (and superconducting) even though all U sites are crystallographically equivalent. An approximately 1% decrease in the ordered moment[38] and pronounced changes in the spin-excitation spectra below $T_c$ confirm that these 5f electrons, though with very different characters, are intimately coupled to each other and to superconductivity.[39] These experiments are among the most definitive in showing that antiferromagnetic spin fluctuations are a viable mechanism for unconventional superconductivity.

The isostructural compound $UNi_2Al_3$ is similar to but also distinctly different from $UPd_2Al_3$. Though antiferromagnetism and superconductivity also coexist in $UNi_2Al_3$ and these two orders are coupled, the ordered moment is much smaller ($\sim 0.2$ $\mu_B$) and is an incommensurate spin-density-wave type.[40] In further contrast to $UPd_2Al_3$, so far there is no clear evidence that the 5f electrons in $UNi_2Al_3$ assume dual roles, possibly because all of the f electrons are more nearly itinerant.[41] The most striking difference, however, is in the nature of superconductivity. Knight shift studies[42] of $UNi_2Al_3$ are consistent with an odd-parity superconducting gap, i.e., S=1 and L=1; whereas, similar measurements on $UPd_2Al_2$ imply an even-parity gap (S=0, L=2) with line nodes.[43] A spin-triplet state (S=1), also found[44] in hexagonal $UPt_3$, typically is expected if Cooper pairing is mediated by ferromagnetic spin fluctuations, as opposed to antiferromagnetic fluctuations that favour spin-singlet (S=0) pairing. [9]

Hybridisation of 5f and ligand electrons creates a narrow, highly correlated electronic band at $E_F$ out of which 5f-band ferromagnetism emerges in $UGe_2$, UIr and URhGe with

Curie temperatures (ordered moments) of 54 K (1.48 $\mu_B$)[23], 46 K (0.5 $\mu_B$)[24], and 9.5 (0.42 $\mu_B$)[27], respectively. Superconductivity appears at atmospheric pressure in URhGe[27] and is induced by pressure in UGe$_2$[23] and UIr[24]. Because their superconductivity develops in proximity to ferromagnetism, it is plausible that superconductivity in these compounds might be unconventional, spin-triplet. Very little is known about the superconductivity in UIr, except that it exists in a very narrow pressure window slightly above 2.5 GPa where a second ferromagnetic transition in UIr appears to extrapolate to T=0.[24] On the other hand, UGe$_2$ and URhGe, which form in similar orthorhombic crystal structures, have been studied somewhat more extensively. Their large normal state specific heat coefficients, $\gamma$ $\approx$120-160mJ/mol-K$^2$, clearly imply that 5f electrons participate in band ferromagnetism, but $\Delta C/\gamma T_c$ in both compounds is only about one-third that expected by BCS theory.[27] This second property suggests either that the sample volume may not completely superconducting, e.g. superconductivity might exist only at the interface of ferromagnetic domains, or that a reduced $\Delta C/\gamma T_c$ could the manifestation of an unavoidable vortex state created by the internal magnetic field associated with ferromagnetic order. Whether superconductivity and ferromagnetism coexist or compete in these compounds is an important issue that requires further investigation, but initial spin-lattice relaxation experiments on polycrystalline UGe$_2$ are consistent with bulk, spin-triplet superconductivity.[45] Finally, we note that pressure studies reveal strikingly different relationships between superconductivity and ferromagnetism in UGe$_2$ and URhGe. The Curie temperature of UGe$_2$ decreases monotonically to T$_C$=0 at a critical pressure of ~1.6 GPa, and superconductivity exists between ~0.9 and ~1.6 GPa.[23] In URhGe, however, superconductivity at atmospheric pressure is completely suppressed with about 3-3.5 GPa applied pressure, even though its T$_C$ increases linearly to over 20 K at 13 GPa.[31]

## 2.2 Plutonium Superconductors

Perhaps the most surprising recent development in actinide superconductivity has been the discovery of superconductivity in isostructural compounds PuCoGa$_5$[28] and PuRhGa$_5$[29]. Not only are these the first Pu-based superconductors, their T$_c$'s are nearly an order of magnitude higher than for other actinides and are comparable to many of the highest T$_c$'s of transition metal compounds. With nearest f-f spacings well beyond the Hill limit and Curie-Weiss-like uniform magnetic susceptibilities, these Pu-based superconductors appear to have localized 5f electrons. Their enhanced Sommerfeld coefficients, $\gamma$ $\approx$70-100 mJ/mol-K$^2$, and BCS-like jump $\Delta C/\gamma T_c$, however, are consistent with bulk superconductivity developing out of a relatively narrow, correlated band of conduction electrons.[28,29,46,47] In these respects, PuCoGa$_5$ and PuRhGa$_5$ are reminiscent of UPd$_2$Al$_3$. A further similarity is revealed in photoemission spectra of PuCoGa$_5$ that are described best by a model in which four of Pu's five 5f electrons are localized and one 5f electron is itinerant, as deduced as well for δ-Pu.[48] This picture of Pu's 5f configuration still leaves open the possibility that superconductivity is conventional, especially given the surprisingly high T$_c$'s. Because Pu is much more radioactive than U, self-heating makes it difficult to study properties of these Pu compounds at temperatures much less than 4 K << T$_c$, where ideally experiments should be made to differentiate between conventional and unconventional superconductivity. Nevertheless, power-laws in specific heat[46] and, most convincingly, in the spin-lattice relaxation rate 1/T$_1$ well below T$_c$ of PuCoGa$_5$[49] and PuRhGa$_5$[50] argue strongly for a superconducting gap with nodes. Combined with Knight-shift measurements that establish spin-singlet pairing in PuCoGa$_5$[49], these power-law dependences are consistent with Cooper pairs having net angular momentum L=2. This pairing state is the same as that in the isostructural heavy-fermion superconductor $CeCoIn_5$, whose $\gamma$ is about ten times larger and $T_c$ about ten times smaller than found in $PuCoGa_5$.[51] Further, above $T_c$, $1/T_1$ of both $CeCoIn_5$ and $PuCoGa_5$ exhibits a temperature dependence that is dominated by low-energy antiferromagnetic spin fluctuations[49], which are favourable to formation of an unconventional superconducting state.

## 2.3  Am Superconductivity

Elemental Am is the only known trans-Pu superconductor. With six well-localized 5f electrons, Hund's rules require that the ground state of $Am^{3+}$ be non-magnetic. Consequently, it is probable that, at atmospheric pressure, phonon-mediated superconductivity arises[30] in a band of weakly correlated non-f electrons, which is inferred from Am's transition-metal-like $\gamma \approx 3$ mJ/mol-$K^2$.[52] Applying pressure to Am induces a cascade of structures with progressively smaller unit-cell volumes and a remarkable increase of $T_c$ from 0.79 K at P=0 to a maximum of 2.3 K at the Am-I/Am-II border near 6 GPa.[53,54] $T_c$ decreases monotonically across the Am-II and Am-III phases, reaches a minimum of ~1.1 K near the Am-III/Am-IV border (~16 GPa) and varies non-monotonically with increasing pressure in Am-IV. Pronounced changes in the resistivity accompany these changes in structure and $T_c$, leading to the suggestion of a pressure-induced localized to delocalised transition of the 5f electrons that plausibly accounts for the non-monotonic $T_c(P)$ in Am-IV.[54] Though much work remains to be done, it seems likely that superconductivity in at least one of these high-pressure phases of Am could be unconventional.

## 3  PERSPECTIVE

For unconventional superconductivity mediated by spin fluctuations, $T_c \sim T_{sf}e^{-1/\lambda}$, where $T_{sf}$ is the characteristic energy scale of spin fluctuations and $\lambda$ is parameter of order unity that characterizes the coupling of Cooper pairs by spin fluctuations. Provided $\lambda$ does not vary significantly among different materials, this analogue of the BCS expression (where $T_{sf}$ is replaced by a characteristic phonon energy) implies an approximately linear relationship between $T_c$ and $T_{sf}$, which is shown in Fig. 2. This correlation argues for a continuum in the pairing mechanism among Ce- and U-based heavy-fermion and the other major family of unconventional superconductors, high-$T_c$ cuprates.[49] Significantly, $PuCoGa_5$ bridges the gulf between these extremes. Qualitatively, this can be understood because $T_{sf}$ reflects the degree to which nearly localized and ligand electrons hybridise, i.e., the bandwidth of strongly correlated itinerant electrons. The relatively large radial extent of 3d-electron wave functions allows greater mixing with ligand wave functions relative to ligand mixing with 4f wave functions, which are confined more closely to the ionic core. Plutonium's 5f electrons are intermediate between 3d and 4f behaviours. To a first approximation, then, $T_c$'s of these superconductors are determined by the relative widths of their highly correlated conduction bands. Though there is scatter among points included in Fig. 2, due in part to different methods used to estimate $T_{sf}$,[9] it is remarkable that the correlation is as good as it is, given the diversity of ground states and crystal structures, but it is interesting to note that those believed to be spin-triplet superconductors lie relatively further away from the straight line. The rich spectrum of physical properties exhibited by actinide superconductors has challenged our understanding of 5f electrons and will continue to exert significant influence on the study of strongly correlated electron phenomena and materials. As with transition-metal physics and conventional superconductivity, the discovery of new actinide superconductors will be important for guiding the development of appropriate theory of both the condensed matter physics of 5f electrons and of unconventional superconductivity. In this regard, exploration of ternary compounds with a tetragonal structure that provides f-f spacing beyond Hill's limits may prove beneficial.

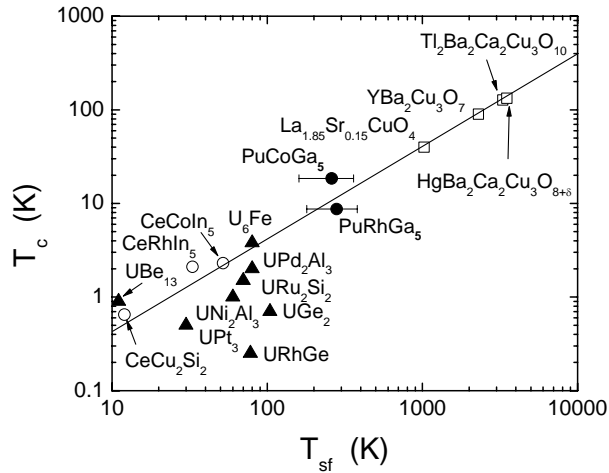

**Figure 2**    $T_c$ versus $T_{sf}$ on logarithmic scales for several actinide, rare-earth and cuprate superconductors. The solid line has unity slope. Adapted after ref. 9.